\documentclass{iopart}

\usepackage[utf8]{inputenc}
\usepackage{graphicx}
\usepackage{units}
\usepackage{color}
\usepackage{bbold}
\usepackage{iopams}
\usepackage{cite}
\usepackage{wrapfig}
\usepackage{comment}
\usepackage{bbold}
\usepackage{enumerate}

\usepackage[pdftex, colorlinks=true, linkcolor=myblue, citecolor=myblue,
urlcolor=myblue]{hyperref}

\definecolor{myblue}{rgb}{0,0,1}

\newcommand{\q}{\mathbf{q}}
\newcommand{\R}{\mathbf{R}}

\begin{document}

\title[Dirac plasmons in bipartite lattices of metallic nanoparticles]{Dirac plasmons in bipartite lattices of metallic nanoparticles}

\author{Thomas Jebb Sturges$^1$, Claire Woollacott$^1$, Guillaume Weick$^2$ and Eros Mariani$^1$}
\address{$^1$Centre for Graphene Science, Department of Physics and Astronomy, University of Exeter, EX4 4QL Exeter, United Kingdom}
\address{$^2$Institut de Physique et Chimie des Mat\'{e}riaux de Strasbourg, Universit\'{e} de
Strasbourg, CNRS UMR 7504, F-67034 Strasbourg, France}

\ead{\href{mailto:Guillaume.Weick@ipcms.unistra.fr}{Guillaume.Weick@ipcms.unistra.fr}
and \href{mailto:E.Mariani@exeter.ac.uk}{E.Mariani@exeter.ac.uk}}

\begin{abstract}
We study theoretically ``graphene-like" plasmonic metamaterials constituted by two-dimensional arrays of metallic nanoparticles, including perfect honeycomb structures with and without inversion symmetry, as well as generic bipartite lattices. The dipolar interactions between localised surface plasmons in different nanoparticles gives rise to collective plasmons that extend over the whole lattice. We study the band structure of collective plasmons and unveil its tunability with the orientation of the dipole moments associated with the localised surface plasmons. Depending on the dipole orientation, we identify a phase diagram of gapless or gapped phases in the collective plasmon dispersion. 
We show that the gapless phases in the phase diagram are characterised by collective plasmons behaving as massless chiral Dirac particles, in analogy with electrons in graphene. When the inversion symmetry of the honeycomb structure is broken, collective plasmons are described as gapped chiral Dirac modes with an energy-dependent Berry phase.
We further relax the geometric symmetry of the honeycomb structure by analysing generic bipartite hexagonal lattices. In this case we study the evolution of the phase diagram and unveil the emergence of a sequence of topological phase transitions when one hexagonal sublattice is progressively shifted with respect to the other.
\end{abstract}


\maketitle

%==================================================================
%==================================================================
%==================================================================
%==================================================================
\section{Introduction}

The way materials interact with light has been of interest, both scientifically and artistically, for millennia. The first lenses and mirrors were created to bend the trajectory of light, leading to a huge variety of practical applications. Among them, focusing light to small regions of space has been extensively used in classical optics to image tiny objects. However, the diffraction limit has seriously hindered our ability to observe microscopic structures with dimensions less than the wavelength of the detecting light \cite{born}. This limitation has been overcome with the use of plasmonic nanostructures \cite{barne03_Nature, maier}, such as isolated metallic nanoparticles \cite{klar98_PRL}. When illuminated by an external radiation, the electrons in the nanoparticle oscillate collectively, forming a localised surface plasmon (LSP) resonance \cite{bertsch, kreibig}. The evanescent field at the surface of the nanoparticle associated with the LSP \cite{born, bohren} produces strong optical field enhancements in the subwavelength region. This phenomenon overcomes the diffraction limit and allows for resolution at the molecular level \cite{kneip97_PRL}.  

Within the field of plasmonics, single or few nanostructures have been the cynosure so far. However, the attention is rapidly shifting to metamaterials constituted by ordered arrays of nanostructures. The periodic arrangement of the nanostructures in the array, as well as the interactions between the LSPs on each nanostructure can be exploited in a variety of novel artificial materials exhibiting properties far beyond those seen in nature, leading to fascinating applications such as electromagnetic invisibility cloaking \cite{leonh06_Science, pendr06_Science, schur06_Science}, perfect lensing \cite{pendr00_PRL, fang05_Science} and slow light \cite{tsakm07_Nature}. The interaction between LSPs on individual nanostructures generates collective plasmonic modes that extend over the whole array \cite{krenn99_PRL, maier02_PRB, felid02_PRB, maier03_NatureMat, sweat05_PRB, polyu11_NL, quint98_OL, brong00_PRB, park04_PRB, weber04_PRB, koend06_PRB, koend09_NL, han09_PRL}. As a consequence the collective plasmons can exhibit a variety of properties that crucially depend on the lattice structure of the metamaterial, and on the microscopic interactions between LSP resonances.

\begin{figure}[tb]
\centerline{\includegraphics[width=\columnwidth]{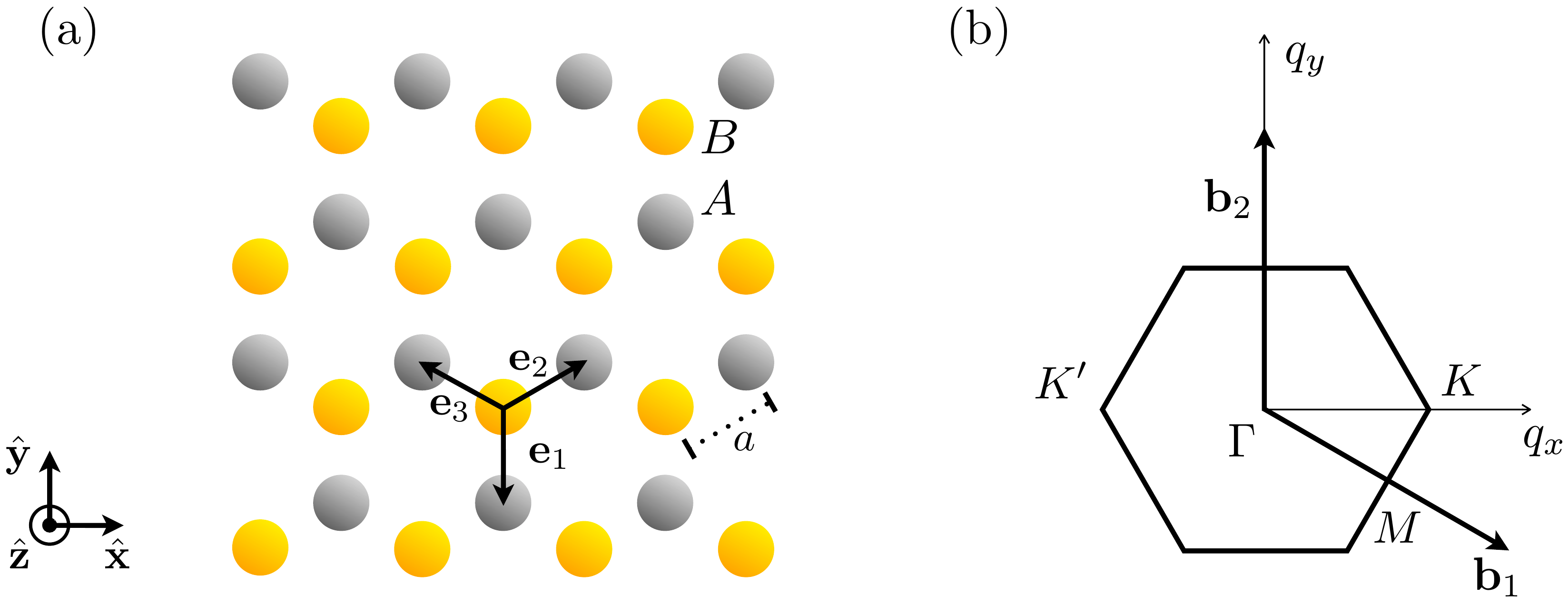}}
\caption{\label{fig:Honeycomb1}%
(a) Honeycomb structure with lattice constant $a$, consisting of two sublattices, labelled $A$ and $B$, connected by the nearest neighbour vectors ${\bf e}_1 = a (0, -1)$, ${\bf e}_2 = a(\frac{\sqrt{3}}{2}, \frac{1}{2})$ and ${\bf e}_3 = a(-\frac{\sqrt{3}}{2}, \frac{1}{2})$. (b) First Brillouin zone with primitive vectors ${\bf b}_1 = \frac{2 \pi}{3 a}(\sqrt{3}, -1)$ and ${\bf b}_2 = \frac{4 \pi}{3 a}(0, 1)$.}
\end{figure}

Among the infinite possible nanostructure arrangements, one structure that has gained considerable attention in the condensed matter community within the last decade is the honeycomb structure of graphene (see figure {\ref{fig:Honeycomb1}}), a two-dimensional (2D) monolayer of carbon atoms \cite{novos04_Science}. The honeycomb structure can be regarded as a special bipartite lattice formed by two inequivalent hexagonal sublattices, denoted by $A$ and $B$, with equal nearest neighbour separation for all carbon-carbon bonds. The hopping of electrons between neighbouring atoms in graphene produces a spectrum characterised by the presence of massless fermionic Dirac quasiparticles \cite{walla47_PR, novos05_Nature, zhang05_Nature, castr09_RMP}. These pseudo-relativistic Dirac fermions have an associated chirality which leads to several of the remarkable properties of graphene, such as a nontrivial Berry phase of $\pi$ responsible for the anomalous quantum Hall effect \cite{novos05_Nature, zhang05_Nature}, as well as the suppression of electronic back-scattering from smooth scatterers \cite{cheia06_PRB}. The latter property is ultimately responsible for the very high electron mobility of graphene samples. 

The majority of the unique electronic properties of graphene stems from the honeycomb structure alone. This feature has been exploited in a variety of ``artificial graphene" systems 
\cite{peleg07_PRL, halda08_PRL, sepkh07_PRA, sepkh08_PRB, zandb10_PRL, bravo12_PNAS, torre12_PRL, tarru12_Nature, weick13_PRL, belle13_PRL, belle13_PRB, polin13_NatureNanotech, jacqm14_PRL} that share the honeycomb lattice structure, albeit in different physical contexts. Among them, in a recent publication \cite{weick13_PRL}, it has been theoretically demonstrated that the properties exhibited by electrons in graphene can be replicated within the context of plasmonic metamaterials. In fact, a 2D honeycomb array of metallic nanoparticles has been shown to support collective plasmons (CPs) that behave as unprecedented chiral massless Dirac bosons. Moreover, the band structure of these collective modes was found to be fully tunable with the polarisation of the incident light that triggers the CPs, a feature which is not available in real graphene devices. Honeycomb plasmonic arrays of metallic nanoparticles thus open interesting perspectives for the realisation of ultrathin tunable metamaterials where the electromagnetic radiation can be transported effectively by chiral pseudo-relativistic Dirac modes.

In this paper we further explore the optical response of bipartite arrays of metallic nanoparticles beyond the artificial graphene plasmonic lattice. 
Firstly, in section \ref{sec:perfect}, we exploit the tunability of the CP band structure and analyse in detail the phase diagram of gapless and gapped phases that emerge while tilting the orientation of the dipole moments associated with LSPs. We show that there are several topologically disconnected regions of the phase diagram which support CPs behaving as massless Dirac quasiparticles. 
In section \ref{sec:broken} we then extend the analysis of the plasmonic array by considering a honeycomb structure with broken inversion symmetry in which the two inequivalent sublattices are made of nanoparticles with different LSP frequencies. Such an approach is relevant for the description of bipartite arrays where the nanoparticles in the two sublattices are either made of different metals, or they are intentionally designed to have different size or shape.
In this case we highlight the emergence of gapped chiral Dirac plasmons characterised by an energy-dependent Berry phase. 
In the last part of the paper (section \ref{sec:deformed}) we explore the interesting scenario of bipartite lattices that break the angular symmetry of the honeycomb array and study the evolution of the aforementioned phase diagram. We show that a sequence of topological phase transitions occurs in the phase diagram while progressively shifting one sublattice with respect to the other. This analysis highlights the existence of massless Dirac plasmons in a vast family of bipartite lattices beyond the honeycomb one, but at the same time it identifies critical values of lattice displacements that lead to the progressive disappearance of massless Dirac plasmons into gapped modes. Our conclusions are finally presented
in section~\ref{sec:ccl}. A short description of the videos accompanying our paper is given in \ref{sec:video}.

%==================================================================
%==================================================================
%==================================================================
%==================================================================
\section{Perfect honeycomb lattice of identical nanoparticles}
\label{sec:perfect}

In this section we set the scene for the rest of the paper by revisiting the properties of an ensemble of identical spherical metallic nanoparticles forming a 2D honeycomb structure \cite{weick13_PRL}, and conducting a thorough analysis of the resultant CPs.

%==================================================================
%==================================================================
%==================================================================
\subsection{Model}

The honeycomb array has a lattice constant $a$ (see figure \ref{fig:Honeycomb1}) and is embedded 
in an (effective) medium with dielectric constant $\epsilon_\mathrm{m}$. Each individual spherical nanoparticle of radius $r$ 
supports an LSP resonance, which can be triggered by an oscillating external electric field, with wavelength $\lambda \gg r$. Under such a condition, the collective electronic excitation associated with the LSP can be modelled as a point dipole with a natural frequency of oscillation $\omega_0$ that depends on the size, shape and structure of the nanoparticles, as well as on their environment \cite{kreibig}. The dipolar LSPs have a dipole moment ${\bf p} = -e N_\mathrm{e} h({\bf R}) \hat{\bf p}$, where $e$ is the electronic charge, $N_\mathrm{e}$ is the number of valence electrons in each nanoparticle, $h(\mathbf{R})$ is the displacement field associated with the electronic centre-of-mass motion that corresponds to the LSP at position ${\bf R}$, and $\hat{\bf p}$ is the dipole unit vector. The LSP can be regarded as a bosonic mode, in particular when the size of the nanoparticle is small enough such that quantum size effects are important \cite{kawab66_JPSJ, yanno92_AP, molin02_PRB, gerch02_PRA, weick05_PRB, weick06_PRB, seoan07_EPJD, weick07_EPJD}.

The nature of the coupling between LSPs in different nanoparticles depends on their size and inter-particle distance. The interactions between the dipolar LSPs lead to collective plasmonic modes
that extend over the whole system. 
Under the condition that $r \lesssim  a/3$, we consider each dipole to interact with its neighbours through dipole-dipole
interactions and only take into account the near field generated by each dipole \cite{brong00_PRB, park04_PRB}. 
Such a quasistatic approximation is reliable if $\lambda\gg a$ and has been shown to qualitatively reproduce the results 
of more sophisticated simulations which include retardation effects \cite{han09_PRL}. Within this approximation, the interaction between 
two dipoles ${\bf p}$ and ${\bf p}'$, located at ${\bf R}$ and ${\bf R}^{\prime}$ respectively, reads
\begin{equation}
\label{eq:V}
\mathcal{V} = \frac{\mathbf{p} \cdot \mathbf{p}^{\prime} - 3 (\mathbf{p} \cdot \mathbf{n}) (\mathbf{p}^{\prime} \cdot \mathbf{n})}{\epsilon_\mathrm{m} |\mathbf{R} - \mathbf{R}^{\prime}|^3}, 
\end{equation}
with $\mathbf{n} = (\mathbf{R} - \mathbf{R}^{\prime})/|\mathbf{R} - \mathbf{R}^{\prime}|$. 
As the CPs will be triggered by an external radiation of a given polarisation, 
we assume that all the dipoles point along the same direction and parametrise their orientations as 
$\hat{\bf p} = \sin\theta ( \sin\varphi \, \hat{\bf x} - \cos\varphi \, \hat{\bf y}) + \cos\theta\, \hat{\bf z}$, 
where $\theta$ is the 
angle between $\hat{\bf p}$ and $\hat{\bf z}$ and $\varphi$ is the angle between the projection of $\hat{\bf p}$ in the 
$xy$-plane and $\mathbf{e}_1$ (see figure \ref{fig:Honeycomb1}). 
Hence, the Hamiltonian reduces to a system of coupled oscillators, 
\begin{equation}
\label{eq:H}
H = H_0 + H_\mathrm{int},
\end{equation}
where the non-interacting term $H_0$ reads as \cite{gerch02_PRA, weick06_PRB}
\begin{equation}
\label{eq:H_0}
H_0 = \sum_{s=A,B} \sum_{{\bf R}_s} \left[ \frac{\Pi_s^2({\bf R}_s)}{2M} + \frac{M}{2} \omega_0^2 h_s^2({\bf R}_s) \right], 
\end{equation}
while the dipole-dipole interaction term is given by \cite{weick13_PRL}
\begin{equation}
\label{eq:H_int}
H_{\mathrm{int}} = \frac{(eN_\mathrm{e})^2}{\epsilon_m a^3} \sum_{\mathbf{R}_B} \sum_{j=1}^3 \mathcal{C}_j h_B({\bf R}_B) h_A({\bf R}_B + {\bf e}_j).
\end{equation}
Here, $s = A,B$ is the sublattice index, $h_s(\mathbf{R}_s)$ is the displacement field associated with the LSP at position $\mathbf{R}_s$ of the lattice, $\Pi_s({\bf R}_s)$ is the conjugated momentum to $h_s({\bf R}_s)$, $M = N_\mathrm{e}m_\mathrm{e}$ is the mass of the displaced valence electrons (with $m_\mathrm{e}$ being the electron mass), and the vectors $\mathbf{e}_j$ ($j=1,2,3$), defined in figure \ref{fig:Honeycomb1}, connect the $A$ and $B$ sublattices. Finally, the coefficients
\begin{equation}
\label{eq:C_j}
\mathcal{C}_j = 1 - 3 \sin^2{\theta}\cos^2{\left(\varphi - \frac{2\pi (j - 1)}{3}  \right)}
\end{equation}
appearing in the Hamiltonian \eref{eq:H_int} depend on the polarisation angles $(\theta,\varphi)$ and hence represent tunable interaction parameters that crucially influence the CP band structure.
Note that in equation \eref{eq:H_int} we only consider the interaction between nearest neighbours, as the effect of interactions beyond nearest neighbours has been shown not to qualitatively change the plasmonic spectrum \cite{weick13_PRL}.

The analogy between our plasmonic structure and the electronic properties of graphene becomes apparent by introducing the bosonic ladder operators
\begin{eqnarray}
a_\R&=&\sqrt{\frac{M\omega_0}{2\hbar}}h_A(\R)+\mathrm{i}\,\frac{\Pi_A(\R)}{\sqrt{2\hbar M 
\omega_0}},
\\
b_\R&=&\sqrt{\frac{M\omega_0}{2\hbar}}h_B(\R)+\mathrm{i}\,\frac{\Pi_B(\R)}{\sqrt{2\hbar M 
\omega_0}},
\end{eqnarray}
which annihilate an LSP on a nanoparticle located at position $\mathbf{R}$ belonging to the $A$ or $B$ sublattice, respectively. These operators obey the commutation relations $[a_{\bf R}^{},a_{{\bf R}^{\prime}}^{\dagger}] = [b_{\bf R}^{},b_{{\bf R}^{\prime}}^{\dagger}] = \delta_{{\bf R},{{\bf R}^{\prime}}}$ and $[a_{{\bf R}}^{},b_{{\bf R}^{\prime}}^{\dagger}] = 0$,  
and give access to the CP dispersion and to the nature of the CP quantum states. The bosonic operators above can be converted to momentum space through $a_{\bf R}= \mathcal{N}^{-1/2}\sum_{\bf q} \exp(\mathrm{i}{\bf q}\cdot {\bf R}) a_{\bf q}$ and  $b_{\bf R}= \mathcal{N}^{-1/2}\sum_{\bf q} \exp(\mathrm{i}{\bf q}\cdot {\bf R}) b_{\bf q}$, with $\mathcal{N}$ the number of unit cells of the honeycomb lattice. Thus the components of the Hamiltonian \eref{eq:H}, equations \eref{eq:H_0} and \eref{eq:H_int}, become
\begin{equation}
\label{eq:H_0_q}
H_0 = \hbar \omega_0\sum_{\bf q} \left(a_{\bf q}^{\dagger} a_{\bf q}^{} + b_{\bf q}^{\dagger} b_{\bf q}^{} \right), 
\end{equation}
\begin{equation}
\label{eq:H_int_q}
H_{\mathrm{int}} = \hbar \Omega \sum_{\bf q} 
\left(
f_\q^{}b_\q^\dagger a_\q^{}+f_\q^* a_\q^\dagger b_\q^{}
\right)
+\hbar \Omega \sum_{\bf q} 
\left(
f_\q^{}b_\q^\dagger a_{-\q}^\dagger+f_\q^* a_{-\q}^{} b_\q^{}
\right),
\end{equation}
respectively. Here we introduced
$\Omega = \omega_0 (r/a)^3 (1 + 2 \epsilon_\mathrm{m}) / 6 \epsilon_\mathrm{m}\ll\omega_0$, while the information on the LSP polarisation is encoded in the function
\begin{equation}
\label{eq:f_q}
f_{\bf q} = \sum_{j=1}^3 \mathcal{C}_j \exp\left(\mathrm{i}\, {\bf q} \cdot {\bf e}_j\right).
\end{equation}

%==================================================================
%==================================================================
%==================================================================
\subsection{Exact diagonalisation}

To find the normal modes of the system we introduce the new bosonic operators 
\begin{equation}
\label{eq:beta}
\beta_{\bf{q}}^\pm = w_\q^\pm a_\q^{} + x_\q^\pm b_\q +
y_\q^\pm a_{-\q}^\dagger + z_\q^\pm b_{-\q}^\dagger
\end{equation}
and impose that the Hamiltonian \eref{eq:H} (with $H_0$ and $H_\mathrm{int}$ given in equations \eref{eq:H_0_q} and \eref{eq:H_int_q}, respectively) is diagonal in this new basis,  
\begin{equation}
\label{eq:H_diag}
H = \sum_\q\left(
\hbar\omega_\q^+{\beta_\q^{+}}^\dagger\beta_\q^{+}
+\hbar\omega_\q^-{\beta_\q^{-}}^\dagger\beta_\q^{-}
\right).
\end{equation}
The Heisenberg equation of motion 
$\left[\beta_\q^\pm , H\right] = \hbar \omega_\q^\pm \beta_\q^\pm$ then 
leads to the eigenvalue problem 
\begin{equation}
\left(\begin{array}{cccc}
\omega_0 & \Omega f_\q & 0 & -\Omega f_\q \\
\Omega f_\q^* & \omega_0 & -\Omega f_\q^* & 0 \\
0 & \Omega f_\q & -\omega_0 & -\Omega f_\q \\
\Omega f_\q^* & 0 & -\Omega f_\q^* & -\omega_0
\end{array}
\right)
\left(
\begin{array}{c}
w_\q^\pm \\
x_\q^\pm \\
y_\q^\pm \\
z_\q^\pm
\end{array}
\right)
=\omega_\q^\pm
\left(
\begin{array}{c}
w_\q^\pm \\
x_\q^\pm \\
y_\q^\pm \\
z_\q^\pm
\end{array}
\right).
\end{equation}
This procedure yields the CP dispersion
\begin{equation}
\label{eq:omega_pm}
\omega_\q^\pm=\omega_0\sqrt{1\pm2\frac{\Omega}{\omega_0}|f_\q|}
\end{equation}
and the coefficients in the coherent superposition \eref{eq:beta}
\begin{equation}
\fl
w_\q^\pm=\frac{\cosh{\vartheta_\q^\pm}}{\sqrt{2}}\frac{f_\q}{|f_\q|},\quad
x_\q^\pm=\pm\frac{\cosh{\vartheta_\q^\pm}}{\sqrt{2}},\quad
y_\q^\pm=-\frac{\sinh{\vartheta_\q^\pm}}{\sqrt{2}}\frac{f_\q}{|f_\q|},\quad
z_\q^\pm=\mp\frac{\sinh{\vartheta_\q^\pm}}{\sqrt{2}}, 
\end{equation}
with
\begin{equation}
\label{eq:cosh}
\cosh{\vartheta_\q^\pm}=\frac{1}{\sqrt{2}}
\left(
\frac{1\pm\Omega |f_\q|/\omega_0}{\sqrt{1\pm2\Omega |f_\q|/\omega_0}}
+1
\right)^{1/2},
\end{equation}
\begin{equation}
\label{eq:sinh}
\sinh{\vartheta_\q^\pm}=\frac{\mp1}{\sqrt{2}}
\left(
\frac{1\pm\Omega |f_\q|/\omega_0}{\sqrt{1\pm2\Omega |f_\q|/\omega_0}}
-1
\right)^{1/2}.
\end{equation}
It is evident by the form of $f_\q$ in equation \eref{eq:f_q} that the dispersion of the two CP branches $\omega_\q^\pm$ crucially depends on the dipole orientation $(\theta, \varphi)$. 
In what follows, we will explore several relevant cases that highlight the tunability of the CP band structures with the LSP polarisation. 

Since $\Omega\ll\omega_0$, the CP dispersion in \eref{eq:omega_pm} can be approximated as 
\begin{equation}
\omega_\q^\pm\simeq\omega_0\pm\Omega|f_\q|, 
\end{equation}
and the coefficients in equations \eref{eq:cosh} and \eref{eq:sinh} reduce to 
$\cosh{\vartheta_\q^\pm}\simeq1$ and $\sinh{\vartheta_\q^\pm}\simeq0$, respectively, so that the Bogoliubov operators in \eref{eq:beta} take the simpler form 
\begin{equation}
\beta_\q^\pm\simeq\frac{1}{\sqrt{2}}\left(\frac{f_\q}{|f_\q|}a_\q\pm b_\q\right).
\end{equation}
Incorporating these expansions in the Hamiltonian \eref{eq:H_diag}, we find that the system is approximately described by the Hamiltonian
\begin{equation}
\label{eq:H_approx}
H\simeq\hbar \omega_0\sum_{\bf q} \left(a_{\bf q}^{\dagger} a_{\bf q}^{} + b_{\bf q}^{\dagger} b_{\bf q}^{} \right)
+\hbar \Omega \sum_{\bf q} 
\left(
f_\q^{}b_\q^\dagger a_\q^{}+f_\q^* a_\q^\dagger b_\q^{}
\right).
\end{equation}
The expression above demonstrates that the non-resonant terms in the interaction Hamiltonian $H_\mathrm{int}$ (cf.\ the third and fourth terms in the right-hand side of equation \eref{eq:H_int_q}) are irrelevant for the description of CPs in the honeycomb array of nanoparticles.

%==================================================================
%==================================================================
%==================================================================
\subsection{Dirac-like collective plasmons}
\label{sec:Dirac_plasmons}

When the LSP polarisation points normal to the plane $(\theta=0)$, the interaction parameters \eref{eq:C_j} are all equal to $1$.
This case is directly analogous to the electronic tight-binding problem in pristine graphene, where the hopping parameters between nearest neighbour carbon atoms are all equal. In fact,
 the spectrum given in \eref{eq:omega_pm} presents two inequivalent Dirac cones \cite{weick13_PRL}, similar to those present in the electronic band structure of graphene \cite{walla47_PR, castr09_RMP}. These Dirac cones occur at a frequency $\omega_{\bf q}^{\pm} = \omega_0$ and are centred at the $K$ and $K'$ points located at $\pm\mathbf{K}_\mathrm{D} = \frac{4 \pi}{3\sqrt{3}a} \left(\pm 1,0\right)$ in the first Brillouin zone (see figure \ref{fig:Honeycomb1}(b)). 
Close to the Dirac points, the function $f_{\bf q}$ defined in \eref{eq:f_q} expands as 
$f_\q \simeq -\frac{3a}{2}(\pm k_x + \mathrm{i} k_y)$, where ${\bf k}$ is the wavevector away 
from the Dirac point (${\bf q} = \pm {\bf K}_\mathrm{D} + {\bf k}$, where $|{\bf k}|\ll|{\bf K}_\mathrm{D}|$), such that the CP
dispersion \eref{eq:omega_pm} is linear and forms a Dirac cone, $\omega_\q^\pm\simeq\omega_0\pm v|\mathbf{k}|$, with group velocity $v=3\Omega a/2$.\footnote{The CP dispersion in the $\theta=0$ case is presented in the video 
\href{http://www.ipcms.unistra.fr/wp-content/uploads/2014/11/tunable_CP_dispersion.mp4}{\texttt{tunable\_CP\_dispersion.mp4}}, 
along with the band structure for a variety of LSP polarisations, see section \ref{sec:PD}.}
We hence find with equation \eref{eq:H_approx} that the system is 
described close to the Dirac points  by the effective Hamiltonian
${H}^{\mathrm{eff}} = \sum_{\bf k} \hat{\Psi}_{\bf k}^{\dagger} \mathcal{H}_{\bf k}^{\mathrm{eff}} \hat{\Psi}_{\bf k}$, where the spinor operator $\hat{\Psi}_{\bf k} = (a_{{\bf k},K}, b_{{\bf k},K}, b_{{\bf k},K^{\prime}}, a_{{\bf k},K^{\prime}})$ and 
\begin{equation}
\label{eq:H_eff}
\mathcal{H}^{\mathrm{eff}}_\mathbf{k} = \hbar \omega_0 \mathbf{1} - \hbar v \tau_z \otimes \boldsymbol {\sigma}\cdot\mathbf{k}.
\end{equation}
Here, $\mathbf{1}$ corresponds to the $4\times4$ identity matrix, $\tau_z$ to the Pauli matrix acting on the valley space ($K/K^{\prime}$), $\boldsymbol{\sigma} = (\sigma_x,\sigma_y)$ is the vector of Pauli matrices acting on the sublattice space ($A/B$), and $a_{\mathbf{k},K(K')}$ and $b_{\mathbf{k},K(K')}$ annihilate plasmons with wavevector $\mathbf{k}$ in the vicinity of the $K$ ($K'$) valley in the $A$ and $B$ sublattices, respectively. 
Up to a global energy shift of $\hbar \omega_0$, equation \eref{eq:H_eff}
corresponds to a massless Dirac Hamiltonian for the CPs and shows that the corresponding spinor eigenstates represent Dirac-like massless bosonic excitations that present similar effects to those of electrons in graphene \cite{weick13_PRL}.

%==================================================================
%==================================================================
%==================================================================
\subsection{Phase diagram for arbitrary localised surface plasmon polarisations}
\label{sec:PD}

We now depart from the specific case of LSPs polarised perpendicular to the plane ($\theta=0$), where there exists a one-to-one correspondence between graphene and our metasurface of metallic nanoparticles. Instead we explore the rich phase diagram of gapless and gapped CP band structures that emerges when tilting the LSP polarisation. 

\begin{figure}[tb]
\centerline{\includegraphics[width=\columnwidth]{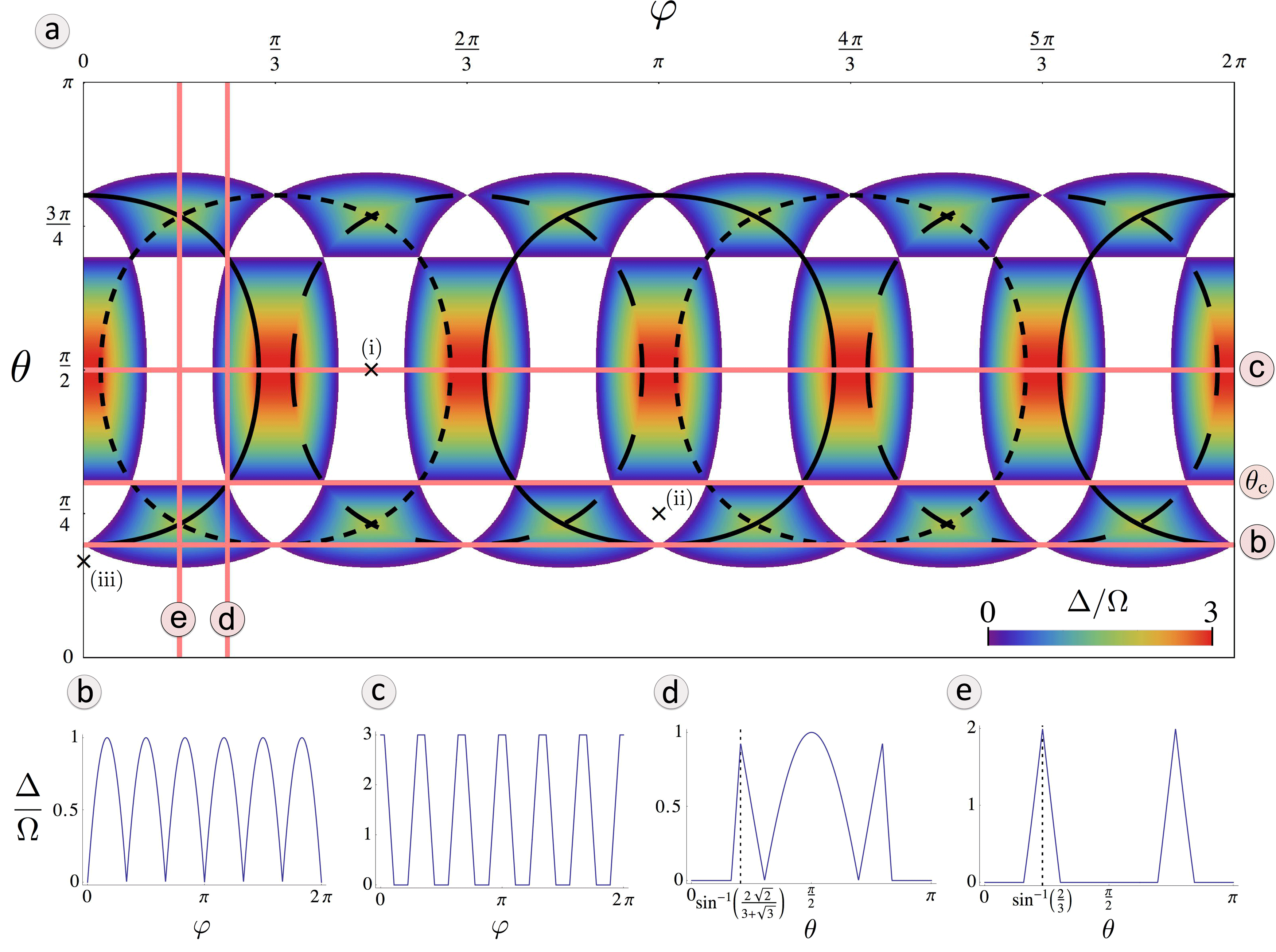}}
\caption{\label{fig:PhaseDiagramHoneycomb}%
(a) Phase diagram for the perfect honeycomb lattice, showing at which polar and azimuthal angles ($\theta$ and $\varphi$, respectively) a gap opens in the spectrum. White indicates a gapless region, while the size of the gap $\Delta$ is given by the colour bar (in units of $\Omega$). The black lines indicate where one of the interaction parameters $\mathcal{C}_j$ is zero 
($\mathcal{C}_{1} = 0$, $\mathcal{C}_2=0$ and $\mathcal{C}_3=0$ along the solid, dashed and dotted lines, respectively). The straight lines labelled (b), (c), (d), and (e) correspond to the cuts shown in the other panels, respectively at (b) $\theta = \arcsin{(1/\sqrt{3})}$, (c) $\theta = \pi/2$, (d) $\varphi = \pi/4$ and (e) $\varphi = \pi/6$. In the figure, $\Omega/\omega_0=0.01$.}
\end{figure}

For an arbitrary polarisation of the LSPs, we can determine if the CP dispersion is gapless by imposing $|f_{\bf q}| = 0$ in \eref{eq:omega_pm}. Together with equation \eref{eq:f_q}, this leads to the condition \cite{weick13_PRL}
\begin{equation}
0 \leqslant \frac{( \mathcal{C}_2 + \mathcal{C}_3 )^2 - \mathcal{C}_1^2}{4 \mathcal{C}_2 \mathcal{C}_3} \leqslant 1
\end{equation}
for having gapless plasmonic modes. This allows us to produce a phase diagram in figure \ref{fig:PhaseDiagramHoneycomb}(a) which shows the polar and azimuthal polarisation angles $\theta$ and $\varphi$, respectively, for which the CP band structure is gapless (white regions) or gapped (coloured regions).
The colour scale in figure \ref{fig:PhaseDiagramHoneycomb}(a) indicates the size of the gap $\Delta$ (in units of the coupling $\Omega$), defined as the difference in energy between the minimum of the upper ($+$) and the maximum of the lower ($-$) band, which for $\Omega/\omega_0\ll1$ reduces to $\Delta\simeq2\Omega\min{\{|f_\q|\}}$.
Figures \ref{fig:PhaseDiagramHoneycomb}(b)-(e) display the size of the gap along the lines (correspondingly labelled) in figure \ref{fig:PhaseDiagramHoneycomb}(a). 
In figure \ref{fig:PhaseDiagramHoneycomb}(a), the black solid, dashed and dotted lines indicate the angles where one of the nearest-neighbour coupling strengths $\mathcal{C}_j$ defined in \eref{eq:C_j} equals zero, reducing the system to a collection of non-interacting 1D chains.
This condition renders the system equivalent to waveguides with a dispersion that is translationally invariant along one direction \cite{weick13_PRL}. As can be seen from figure \ref{fig:PhaseDiagramHoneycomb}(a), there are also points where two of these lines intersect, signalling that two nearest-neighbour bonds are ``cut" and the system can effectively be described as isolated dimers, leading to flat bands with no wavevector dependence. The CP dispersions that arise upon continuously tuning the polarisation angle along these lines is shown in the video 
\href{http://www.ipcms.unistra.fr/wp-content/uploads/2014/11/tunable_CP_dispersion.mp4}{\texttt{tunable\_CP\_dispersion.mp4}}.

\begin{figure}[tb]
\centerline{\includegraphics[width=\columnwidth]{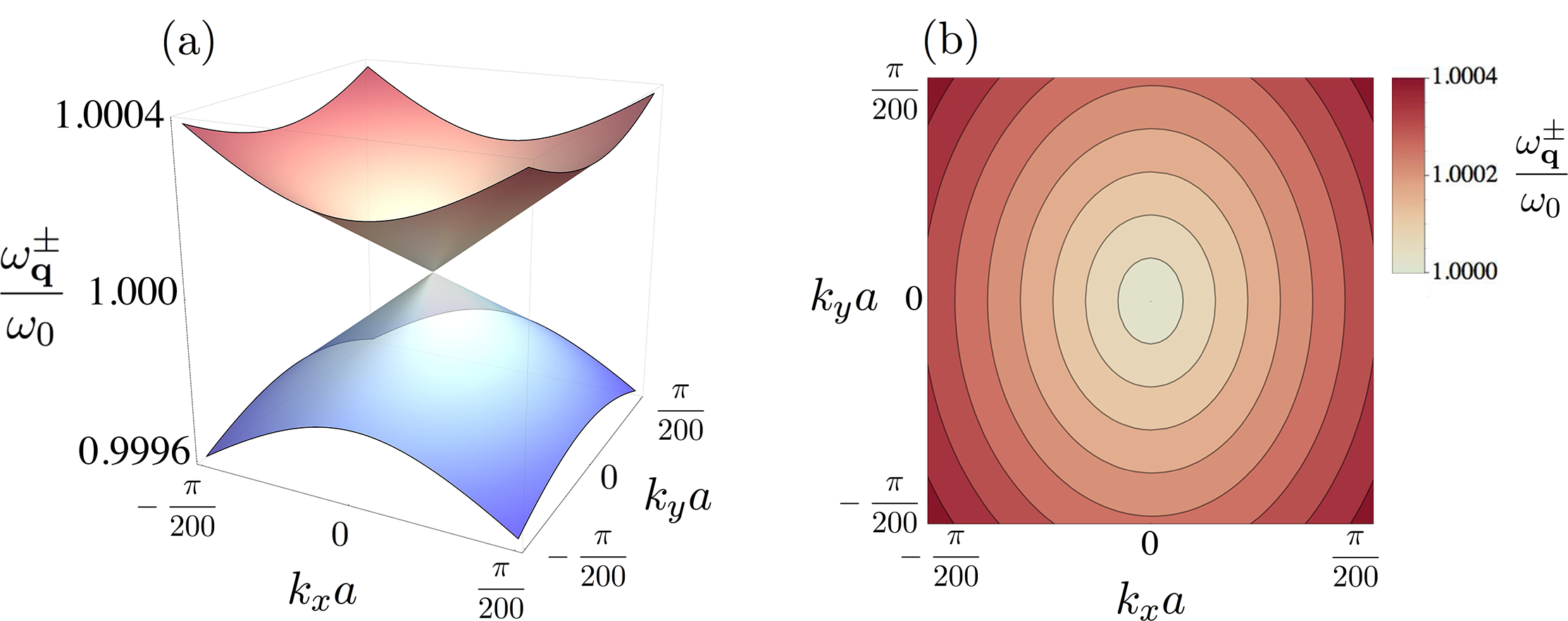}}
\caption{ \label{fig:EllipticalCones}
(a) Collective plasmon dispersion for in-plane polarisation $(\theta, \varphi)=(\pi/2, \pi/2)$ in the vicinity of the Dirac point located at $\mathbf{K}_\mathrm{D}= \frac{2}{\sqrt{3}a}\arccos{(\frac{2}{5})}(1, 0)$. (b) Corresponding isoenergetic lines of the upper $\omega_\q^+$ branch.
In the figure, $\Omega/\omega_0=0.01$ and $\mathbf{k}=\q-\mathbf{K}_\mathrm{D}$.}
\end{figure}

Figure {\ref{fig:PhaseDiagramHoneycomb}}(a) shows that there are many polarisation angles where the spectrum is gapless (see the white regions in the figure), indicating the possibility of Dirac physics. In what follows, we predict Dirac physics in these gapless regions, even in pockets of the phase diagram topologically disconnected from the ``trivial" graphene-like case for which $\theta=0$ (see section \ref{sec:Dirac_plasmons}). To explore this, we expand the Hamiltonian in these regions. We have investigated three different gapless positions: (i) $(\theta, \varphi) = ({\pi}/{2},{\pi}/{2})$, (ii) $(\theta, \varphi) = ({\pi}/4, \pi)$, and (iii) $(\theta, \varphi) = ({\pi}/{6}, 0)$, as
indicated in figure \ref{fig:PhaseDiagramHoneycomb}(a). 
In all three cases there are gapless modes with two cones centred at 
the two inequivalent Dirac points $\pm\mathbf{K}_\mathrm{D}$.
In cases (i) and (ii), these are located at 
$\pm\mathbf{K}_\mathrm{D}= \frac{2}{\sqrt{3}a}\arccos{(\frac{2}{5})}(\pm1, 0)$, and for case (iii), 
$\pm\mathbf{K}_\mathrm{D}= \frac{2}{\sqrt{3}a}\arccos{(-\frac{2}{13})}(\pm 1, 0)$. 
Close to the two inequivalent Dirac points, the function $f_{\bf q}$ of equation \eref{eq:f_q} expands as 
(i) $f_{\bf q} \simeq \frac{3a}{2} (\pm \frac{\sqrt{7}}{2} k_x - \mathrm{i} k_y)$, 
(ii) $f_{\bf q} \simeq \frac{3a}{4} (\mp \frac{\sqrt{7}}{2} k_x + \mathrm{i} k_y)$, 
and 
(iii) $f_{\bf q} \simeq \frac{-3a}{8} ( \pm \frac{\sqrt{55}}{2} k_x + \mathrm{i} k_y)$, 
where ${\bf q} = \pm {\bf K}_\mathrm{D} + {\bf k}$, with $|\mathbf{k}|\ll|\mathbf{K}_\mathrm{D}|$. 
These expansions show that the CP dispersion \eref{eq:omega_pm} forms \textit{elliptical} cones in the vicinity of the Dirac points above, since the magnitude of the $k_x$ and $k_y$ components of $f_\q$ are not equal, unlike in the purely out-of-plane polarisation case (see section \ref{sec:Dirac_plasmons}). This is illustrated in figure \ref{fig:EllipticalCones} for case (i).
Moreover, by expanding equation \eref{eq:H_diag} in the vicinity of the Dirac points, an effective Hamiltonian can be identified for each case that adequately describes the CPs and, up to a global energy shift of $\hbar \omega_0$,  corresponds to a massless Dirac Hamiltonian.
Thus, Dirac-like physics can be recovered in other gapless regions away from the case where the polarisation points normal to the plane ($\theta=0$).

\begin{figure}[tb]
\centerline{\includegraphics[width=0.63\columnwidth, trim= 0cm 0cm 0cm 0cm, clip=true]{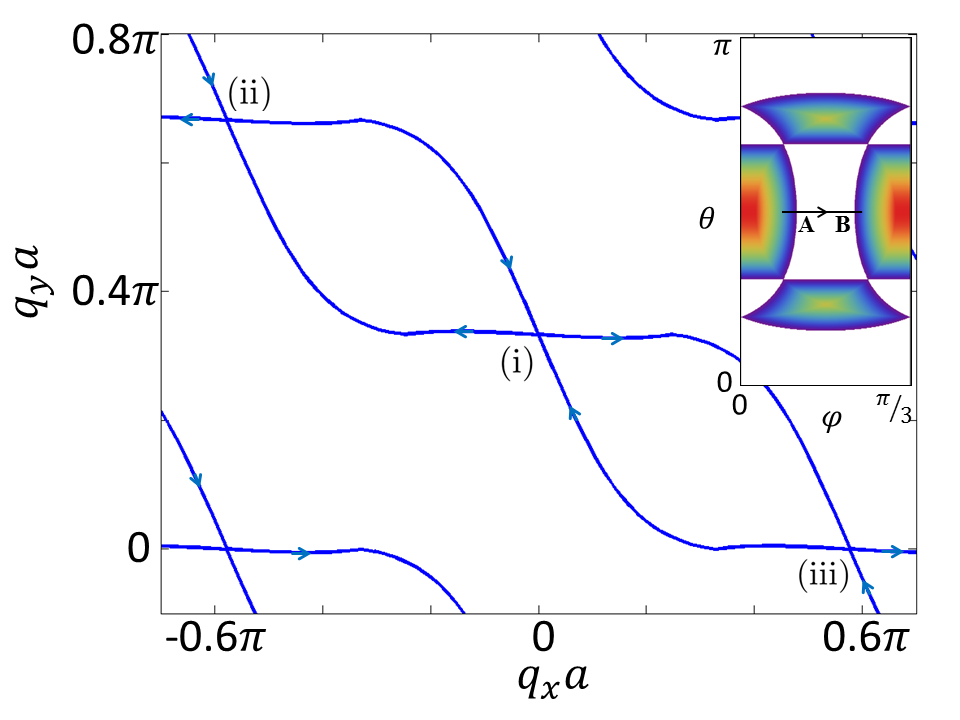}}
\caption{Positions in momentum space of the Dirac points as the LSP polarisation is changed from 
$\varphi\simeq \pi/10$ to $\varphi\simeq \pi/5$, while keeping $\theta=\pi/2$ constant (purely in-plane polarisation), see points A and B in the inset, respectively.
The arrows in the figure indicate the direction of the motion of the Dirac points while $\varphi$ increases.
Inset: section of the phase diagram from figure \ref{fig:PhaseDiagramHoneycomb}(a) showing the polarisation angles considered here. In the figure, $\Omega/\omega_0=0.01$.}
\label{fig:DiracPointMovements}
\end{figure}

The emergence of a gap in the CP spectrum occurs by the annihilation of Dirac points with opposite Berry phases while varying the polarisation direction. This is illustrated in figure \ref{fig:DiracPointMovements} which shows the position of the Dirac points when continuously varying the polarisation angle from $\varphi\simeq\pi/10$ to $\varphi\simeq\pi/5$ while keeping $\theta=\pi/2$ constant (purely in-plane polarisation). These two extrema are denoted by A and B in the inset of the figure, which reproduces part of the phase diagram of figure \ref{fig:PhaseDiagramHoneycomb}(a). 
As can be seen from figure \ref{fig:DiracPointMovements}, the Dirac points move in momentum space as $\varphi$ is increased (see the arrows in the figure that indicate the direction of the motion). At the angles where a gap opens,  the two inequivalent Dirac points merge and annihilate each other. Hence in the gapped regions there are no longer two distinct valleys created by the two sublattices. At the angle labelled A in the inset of figure \ref{fig:DiracPointMovements} ($\theta = \pi/2$, $\varphi \simeq \pi/10$), the gap in the spectrum closes and the two Dirac points in each Brillouin zone appear at the same position, for example two at point (i), two at point (ii) and two at point (iii) in figure \ref{fig:DiracPointMovements}. At the angle labelled B in the inset of the figure ($\theta = \pi/2$, $\varphi \simeq \pi/5$) the two Dirac points in each Brillouin zone coalesce to open a gap in the spectrum, for example one that started at point (ii) and one that started at point (iii) coalesce at point (i). The evolution of this coalescence is shown in video 
\href{http://www.ipcms.unistra.fr/wp-content/uploads/2014/11/tunable_CP_dispersion.mp4}{\texttt{tunable\_CP\_dispersion.mp4}}.

\begin{figure}[tb]
\centerline{\includegraphics[width=\columnwidth]{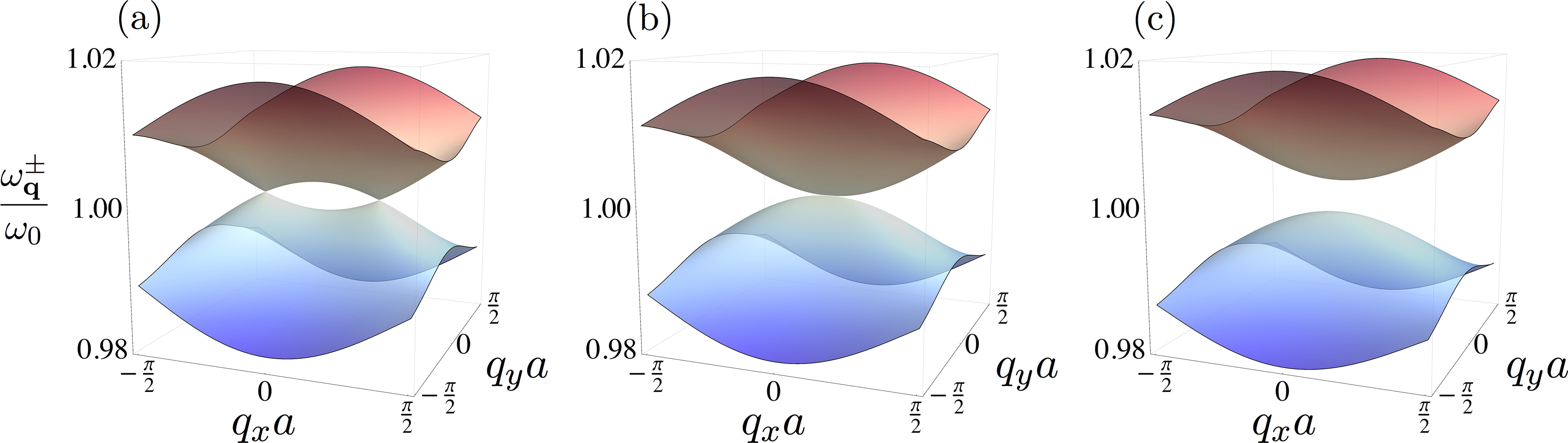}}
\caption{Collective plasmon dispersion for (a) $\theta=\frac{52}{180}\pi$,  
(b) $\theta=\theta_\mathrm{c}\simeq\frac{54.7}{180}\pi$, and (c) $\theta=\frac{58}{180}\pi$, with $\varphi=0$ and $\Omega/\omega_0=0.01$.}
\label{fig:DiracPointMerging}
\end{figure}

The phase diagram in figure {\ref{fig:PhaseDiagramHoneycomb}}(a) shows a symmetry line at $\theta_\mathrm{c} = \arcsin{(\sqrt{2/3})} \simeq \frac{54.7}{180}\pi$, where the dispersion for a given $\varphi$ either changes from gapped to gapless or vice versa when $\theta$ is increased. In figure \ref{fig:DiracPointMerging} we investigate for $\varphi=0$ the dispersion at $\theta$ values just below $\theta_\mathrm{c}$ (figure \ref{fig:DiracPointMerging}(a)), exactly at $\theta=\theta_\mathrm{c}$ (figure \ref{fig:DiracPointMerging}(b)), and just above $\theta_\mathrm{c}$ (figure \ref{fig:DiracPointMerging}(c)). It is evident from the figure that two inequivalent Dirac cones annihilate at $\q=0$ when $\theta=\theta_\mathrm{c}$, subsequently opening a gap and forming two paraboloids for $\theta>\theta_\mathrm{c}$. Indeed, this becomes transparent when expanding \eref{eq:f_q} in the vicinity of $\q=0$. Introducing $\theta=\theta_\mathrm{c}+\delta\theta$ with $0\leqslant\delta\theta\ll\theta_\mathrm{c}$, we find
\begin{eqnarray}
\label{eq:f_q_zero}
f_\q&\simeq&-3\left[\sqrt{2}\,\delta\theta+\frac{1}{8}(q_xa)^2\right]+\mathrm{i}\frac{3}{2}q_ya,
\\
|f_\q|&\simeq&3\sqrt{2}\,\delta\theta+\frac{3\sqrt{2}}{16}\left[\sqrt{2}(q_xa)^2+\frac{1}{\delta\theta}(q_ya)^2\right].
\end{eqnarray}
Thus, any finite positive $\delta\theta$ leads to the formation of a gap in the CP spectrum at $\q=0$ (see equation \eref{eq:omega_pm}). In the vicinity of this point, the bands are parabolic with two inequivalent effective masses whose ratio is controlled by $\delta\theta$ (see figure \ref{fig:DiracPointMerging}(c)). 
In this case, the gap in the spectrum is due to the non-vanishing $f_\q$ at $\q=0$ (see equation \eref{eq:f_q_zero}) 
and does not require breaking of the symmetry between the two sublattices (see section \ref{sec:broken}).

The phenomenon described in figures \ref{fig:DiracPointMovements} and \ref{fig:DiracPointMerging} 
 is reminiscent of Dirac point merging in graphene that has been predicted to occur with large mechanical deformations of the lattice \cite{monta09_PRB, monta09_EPJB}. While these deformations seem to be impossible to reach experimentally in real graphene, other systems 
 such as cold atoms in optical lattices \cite{tarru12_Nature} or microwaves in artificial deformed honeycomb structures \cite{belle13_PRL, belle13_PRB} have been realised and Dirac point merging has been observed. The experimental feasibility of our proposal to observe this phenomenon only requires the external light polarisation to be varied.

%==================================================================
%==================================================================
%==================================================================
%==================================================================
\section{Honeycomb structure with broken inversion symmetry}
\label{sec:broken}

In this section we consider a perfect honeycomb array of spherical metallic nanoparticles with broken inversion symmetry by envisaging that the LSPs on the $A$ and $B$ sublattices have inequivalent resonance frequencies, $\omega_A$ and $\omega_B$, respectively. This could be realised experimentally by either manufacturing the two sublattices out of different materials or by constructing them with different sizes \cite{kreibig}. For our quasistatic approximation of point-like interacting dipoles to hold we still assume the radii ($r_A$ and $r_B$) of the particles in the two sublattices to be much smaller than $\lambda$.
The analysis in section \ref{sec:perfect} applies to this case as well with only minor changes: (i) the natural LSP frequency $\omega_0^{}$ is replaced by $\omega_A$ and $\omega_B$ in the two sublattices, and (ii) the interaction coefficient $\Omega$ is replaced by $\tilde{\Omega}=(\omega_{A}^{}\omega_{B}^{}/\omega^{2}_{0})^{1/2}_{}(r_{A}^{}r_{B}^{}/r^{2}_{})^{3/2}_{}\, \Omega$.
As a result, the dispersions of the two CP branches now read
\begin{equation}
\omega_{\bf q}^{\pm} = \sqrt{\frac{\omega_A^2 + \omega_B^2}{2} 
\pm \sqrt{\left( \frac{\omega_A^2 - \omega_B^2}{2} \right)^2 + 4 \omega_A \omega_B\tilde{\Omega}^2 |f_{\bf q}|^2 }} ,
\label{equ:FinalDispersion}
\end{equation}
where $f_\q$ is defined in equation \eref{eq:f_q}.

\begin{figure}[tb]
\centerline{\includegraphics[width=\columnwidth]{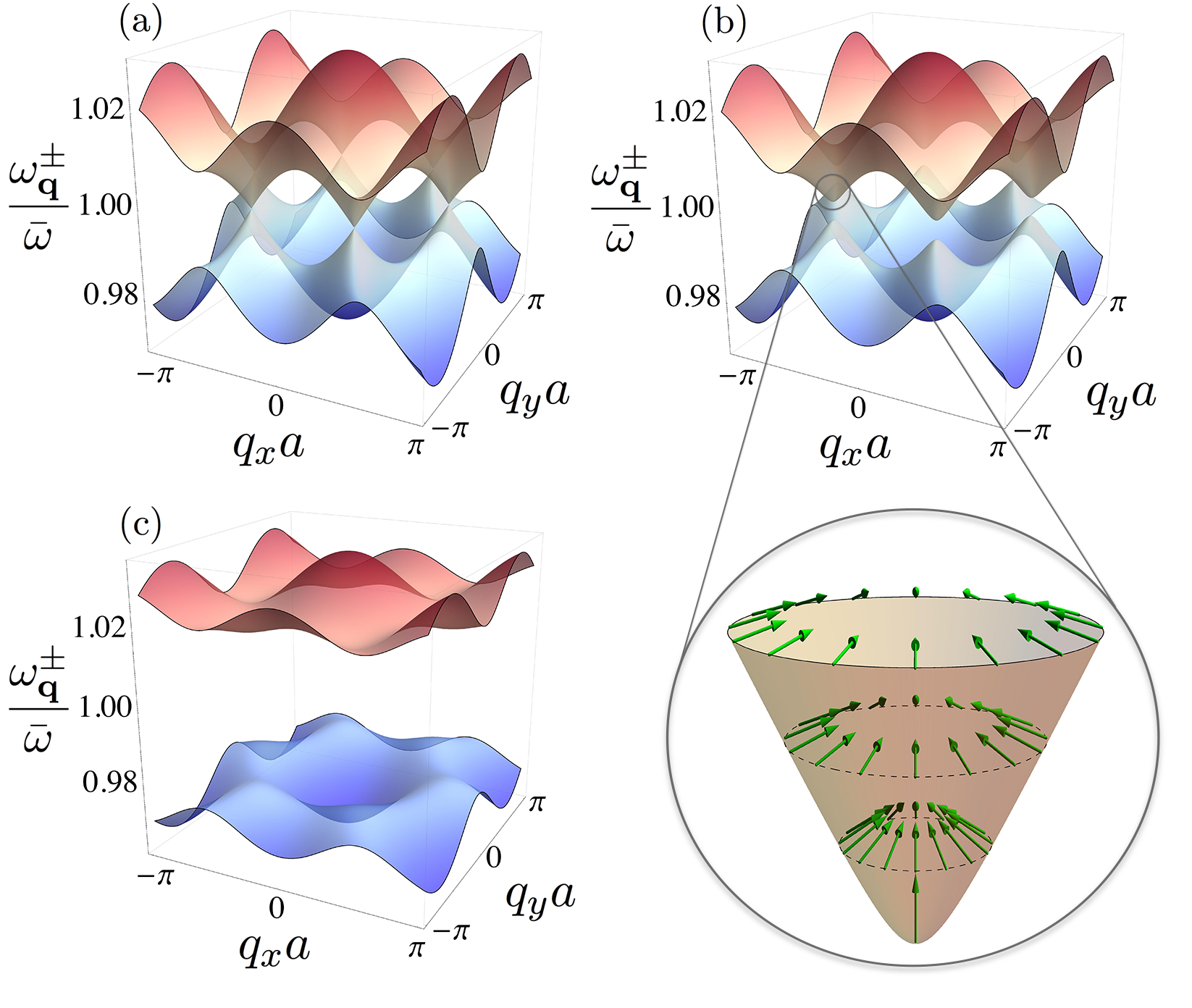}}
\caption{{Collective plasmon dispersions from equation {\eref{equ:FinalDispersion}} with LSP polarisation 
normal to the plane ($\theta=0$), for (a) $\delta\omega/\bar \omega=0$, (b) $\delta\omega/\bar\omega=0.004$ and (c) $\delta\omega/\bar\omega=0.04$. 
The zoom in panel (b) shows the upper collective plasmon branch in the vicinity of the $K$ point in the first Brillouin zone, while the arrows depict the unit vector $\mathbf{V}^+_\mathbf{k}$ defined in the text. In the figure, $\tilde\Omega/\bar\omega=0.01$.}}
\label{fig:Dispersions}
\end{figure}

\begin{figure}[tb]
\centerline{\includegraphics[width=0.63\textwidth, trim=0cm 0cm 0cm 0cm, clip=true]{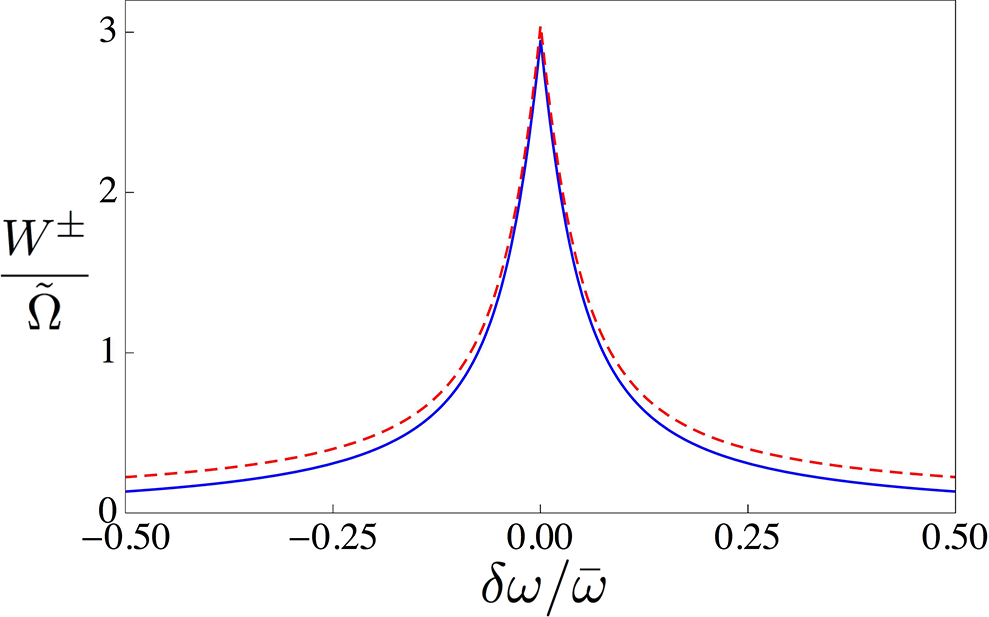}}
\caption{Widths $W^{+}_{}$ and $W^{-}_{}$ of the upper and lower plasmonic bands (blue solid and red dashed lines, respectively) in units of $\tilde{\Omega}$ as a function of $\delta\omega/\bar{\omega}$. 
In the figure, $\tilde\Omega/\bar\omega =0.01$.}
        \label{fig:DiffParticleWidthsFuncofBsize}
\end{figure}

It is convenient to express the two LSP frequencies as
$\omega_A = \bar{\omega} + \delta \omega/2$ and $\omega_B = \bar{\omega} - \delta \omega/2$,
with $\bar{\omega}$ the average frequency and $\delta \omega$ the difference in frequency.\footnote{For definiteness, from now on we assume that $\delta\omega\geqslant0$, without loss of generality.}
 In figure \ref{fig:Dispersions} we plot the dispersion \eref{equ:FinalDispersion} in the special case where the polarisation points perpendicular to the 2D plane ($\theta =0$), for $\delta\omega /\bar{\omega} = 0$ (figure \ref{fig:Dispersions}(a)), $\delta\omega /\bar{\omega} = 0.004$ (figure \ref{fig:Dispersions}(b)) and $\delta\omega /\bar{\omega} = 0.04$ (figure \ref{fig:Dispersions}(c)). As can be seen from the figure, 
any finite difference in the LSP frequencies ($\delta\omega\neq 0$) introduces an asymmetry into the system that corresponds to a gap of size $\delta\omega$ opening in the spectrum while the extrema of the two bands still occur at the $K$ and $K^{\prime}_{}$ points in the Brillouin zone. Notice also that the symmetry-breaking term $\delta\omega$ affects the bandwidth of each plasmonic subband, defined as $W^{\pm}_{}=\max{\{\omega^{\pm}_{\q}\}}-\min{\{\omega^{\pm}_{\q}\}}$, as shown in figure {\ref{fig:DiffParticleWidthsFuncofBsize}}. In particular, increasing $\delta\omega$ leads to an algebraic decrease of both bandwidths. These effects of the symmetry-breaking term $\delta\omega$ are further examplified in the video 
\href{http://www.ipcms.unistra.fr/wp-content/uploads/2014/11/broken_inversion_symmetry.mp4}{\texttt{broken\_inversion\_symmetry.mp4}}.

In the vicinity of the Dirac points, the system is effectively described by the Hamiltonian for gapped quasiparticles
\begin{equation}
\label{eq:H_DiracGap}
\mathcal{H}^{\mathrm{eff}}_\mathbf{k} = \hbar \bar{\omega} \mathbf{1} + \frac{\hbar\delta\omega}{2} \tau_{z}^{}\otimes\sigma_{z}^{} - \hbar \tilde{v} \tau_z \otimes \boldsymbol {\sigma}\cdot\mathbf{k},
\end{equation}
with eigenvalues $\omega_{{\bf k}}^{\pm}= \bar{\omega}\pm \sqrt{\tilde{v}_{}^{2}|{\bf k}|^{2}_{}+(\delta\omega/2)^{2}_{}}$, where $\tilde{v}=3\tilde{\Omega}a/2$. In contrast to the analysis in section \ref{sec:PD}, here the dispersion is gapped for any value of wavevector and the quasiparticles acquire a finite effective mass $m^{*}_{}=\hbar\delta\omega/2\tilde{v}^{2}_{}$ due to the inversion symmetry breaking term $\delta\omega\neq 0$. This scenario is reminiscent of the electronic dispersion in hexagonal boron nitride \cite{doni69_NuovoCimento} and in transition metal dichalcogenides, such as MoS$_2$ 
\cite{xiao12_PRL, korma13_PRB, cappe13_PRB}.

Besides the global frequency shift $\bar{\omega}$, the effective Hamiltonian \eref{eq:H_DiracGap} in each valley evolves continuously between a purely out-of-plane Zeeman term $(\hbar\delta\omega/2)\sigma_{z}^{}$ (when $\tilde{v}|{\bf k}|\ll \delta\omega$) and a 2D massless Dirac Hamiltonian $-\hbar\tilde{v}\boldsymbol{\sigma}\cdot\mathbf{k}$ (when $\tilde{v}|{\bf k}|\gg \delta\omega$). The eigenstates of the Hamiltonian in the $K$ valley thus correspond to the normalised vector in the Bloch sphere ${\bf V}_{{\bf k}}^{\pm}=\pm[\tilde{v}_{}^{2}|{\bf k}|^{2}_{}+(\delta\omega/2)^{2}_{}]^{-1/2}_{}(-\tilde{v}k_{x}^{},-\tilde{v}k_{y}^{},\delta\omega/2)$. This allows us to calculate the Berry phase of the pseudo-spin $s=1/2$ Dirac quasiparticles described by \eref{eq:H_DiracGap} as $\phi_{{\rm B}}^{}=s\Omega_{{\rm B}}^{}$, with $\Omega_{{\rm B}}^{}$ the solid angle enclosed by the Bloch-sphere vector ${\bf V}_{{\bf k}}^{\pm}$ while the state $|{\bf k}\rangle$ is transported anticlockwise in a closed loop around the Dirac point in the 2D wavevector space \cite{berry84_PRSA, sepkh08_PRB, sakurai}.
Depending on the ratio $\tilde{v}|{\bf k}|/\delta\omega$, and thus on the quasiparticle energy, the solid angle $\Omega_{{\rm B}}^{}$ changes (see the zoom in figure \ref{fig:Dispersions}(b)). This ranges from $\Omega_{{\rm B}}^{}=0$ at the band gap edges ($\omega=\bar{\omega}\pm \delta\omega/2$ where ${\bf V}_{{\bf k=0}}^{\pm}=\pm \hat{\mathbf{z}}$) to $\Omega_{{\rm B}}^{}\simeq 2\pi$ if $\tilde{v}|{\bf k}|\gg \delta\omega/2$ (where ${\bf V}_{{\bf k}}^{\pm}$ rotates in the $x$-$y$ plane), the latter case being analogous to that found for electrons in monolayer graphene \cite{novos05_Nature, zhang05_Nature}.

Interestingly, we find that CPs in honeycomb arrays with broken inversion symmetry are naturally described as gapped Dirac quasiparticles with an energy-dependent Berry phase (defined up to integer multiples of $2\pi$)
\begin{equation}
\label{eq:Berry_energy}
\phi_{{\rm B}}^{}(\omega)= \pi \frac{|\omega-\bar{\omega}| - \delta\omega/2}{|\omega-\bar{\omega}|}
\end{equation}
in the $K$ valley if $|\omega-\bar{\omega}|\geqslant \delta\omega/2$. A similar analysis yields an equal and opposite Berry phase for the $K'$ valley.

The explicit connection between pseudo spin and orbital degrees of freedom evident in the effective Hamiltonian \eref{eq:H_DiracGap} is at the very heart of several properties exhibited by quasiparticles in real and artificial graphene systems, such as the suppression of elastic backscattering from smooth impurity potentials \cite{cheia06_PRB}, weak antilocalisation \cite{wu07_PRL, tikho09_PRL} as well as the anomalous quantum Hall effect \cite{novos05_Nature, zhang05_Nature}. In a forthcoming publication \cite{wooll_unpublished} we will analyse in detail the consequences of the energy-dependent Berry phase \eref{eq:Berry_energy} in quantum transport of gapped Dirac quasiparticles.

%==================================================================
%==================================================================
%==================================================================
%==================================================================
\section{Deformed bipartite lattices}
\label{sec:deformed}

\begin{figure}[tb]
\centerline{\includegraphics[width=0.63\columnwidth]{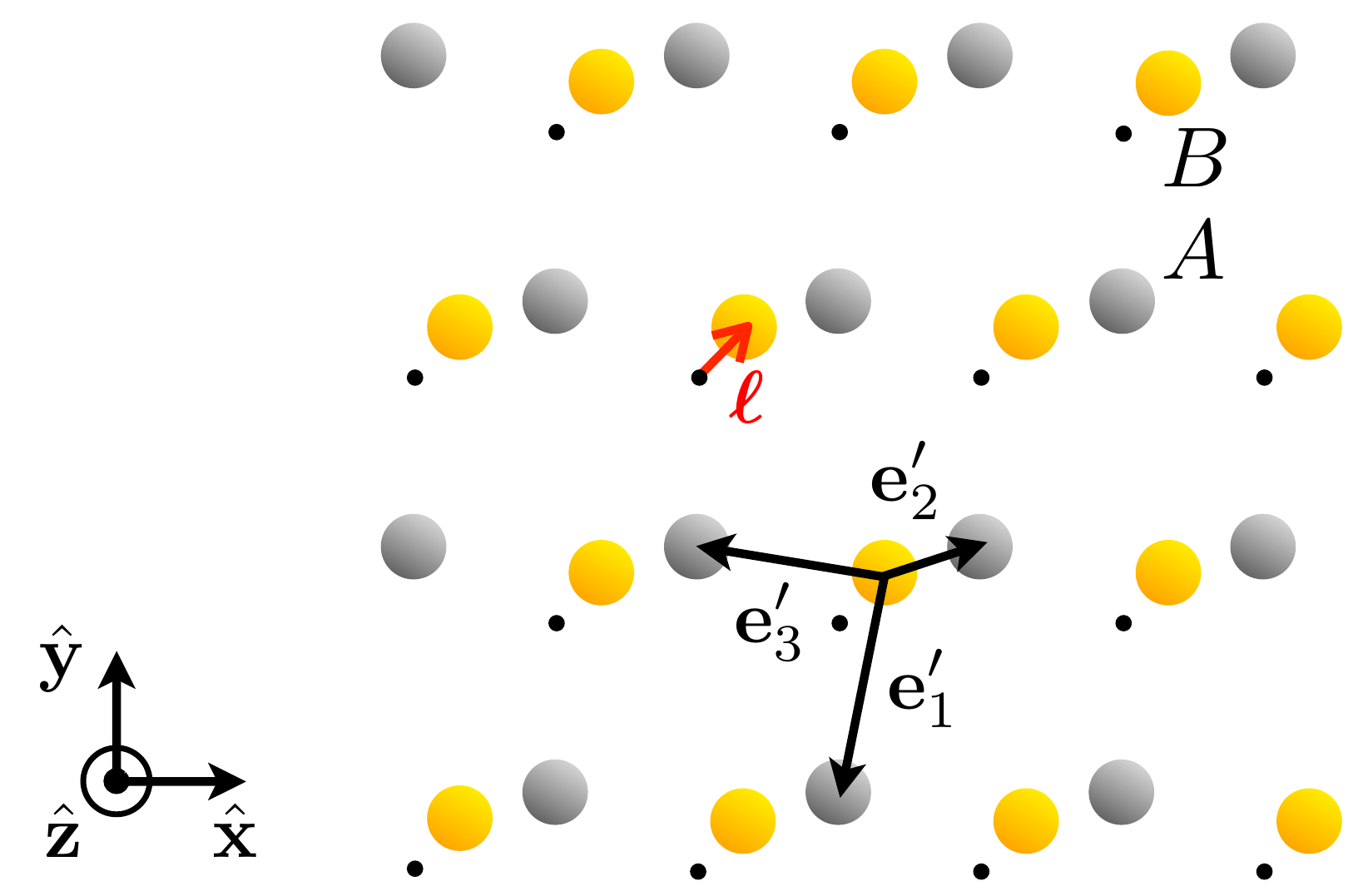}}
\caption{Generic hexagonal bipartite lattice, obtained by displacing the sublattice $B$ of the honeycomb structure by a vector $\boldsymbol{\ell}$. The black dots indicate the original position of the $B$ lattice sites, when $\boldsymbol{\ell}=(0, 0)$, as in figure \ref{fig:Honeycomb1}(a).}
\label{fig:deformed_lattice}
\end{figure}

The four-component wavefunction of the quasiparticles, along with the effective massless Dirac Hamiltonian in real and artificial graphene stem from the bipartite nature of the 2D lattice as well as from the time-reversal symmetric and parity-invariant nature of the system \cite{manes07_PRB}. In particular, the honeycomb structure is a special case of the bipartite hexagonal lattice boasting the additional invariance under rotation by integer multiples of $2\pi/3$. 
In this section we explore the band structure and the stability of the phase diagram of gapless and gapped spectra for CPs in generic hexagonal bipartite lattices obtained by rigidly shifting one sublattice with respect to the other (see figure \ref{fig:deformed_lattice}). 
Our analysis highlights the possibility to design a variety of tunable plasmonic metamaterials with the desired polarisation-dependent optical response, along with the additional benefit of supporting Dirac CPs.

We investigate the properties of 2D arrays of identical spherical metallic nanoparticles with LSP frequency $\omega_0$ sharing the same hexagonal Bravais lattice as the honeycomb structure considered so far, but with an arbitrary position of the second basis nanoparticle. 
Thus the lattice can still be considered as made of two hexagonal sublattices $A$ and $B$, but with the $B$ sublattice shifted by a vector $\boldsymbol{\ell}$ such that the new nearest-neighbour vectors now become ${\bf e}_j^{\prime} = {\bf e}_j - \boldsymbol{\ell}$ (see figures \ref{fig:Honeycomb1}(a) and \ref{fig:deformed_lattice}). This rigid shift breaks the three-fold rotational symmetry of the original honeycomb case. The main questions then to be addressed concern the very existence of Dirac CPs and the fate of the phase diagram while progressively distorting the original lattice.

The mathematical procedure to obtain the Hamiltonian and the band structure for CPs in the general bipartite hexagonal system is a straightforward extension of that presented in section \ref{sec:perfect}. The replacement of the new nearest neighbour vectors ${\bf e}_j^{\prime}$ in the Hamiltonian \eref{eq:H_int} leads to the dispersion relation
\begin{equation}
\omega_{\bf q}^{\pm} = \omega_0 \sqrt{1 \pm 2 \frac{\Omega}{\omega_0} |f_{\bf q}^{\prime}|}, 
\label{equ:sect2dispersion}
\end{equation}
with
\begin{equation}
\label{eq:f_q_prime}
f_{\bf q}^{\prime} = \sum_{j=1}^{3} \mathcal{C}_j^{\prime} e^{\mathrm{i} {\bf q} \cdot {\bf e}_j^{\prime}}.
\label{functionG}
\end{equation}
Here, $\mathcal{C}_j^{\prime} = (a/|{\bf e}_j^{\prime}|)^3(1 - 3 \sin^2 \theta \cos^2 \varphi_j)$
is a generalisation of $\mathcal{C}_j$ defined in equation \eref{eq:C_j}, with 
$\varphi_j = \varphi + \arctan(e_{j,x}^{\prime} / e_{j,y}^{\prime}) $ the angle between ${\bf e}_j^{\prime}$ and the projection of the LSP polarisation $\hat{\bf p}$ into the $x$-$y$ plane.
Unlike in the honeycomb case, the lengths of the nearest neighbour vectors are not all equivalent, hence the extra term $(a / |{\bf e}_j^{\prime}|)^3$ in $\mathcal{C}_{j}^{\prime}$. 

\begin{figure}[tb]
\centerline{\includegraphics[width=\columnwidth, trim=0cm 0cm 0cm 0cm, clip=true]{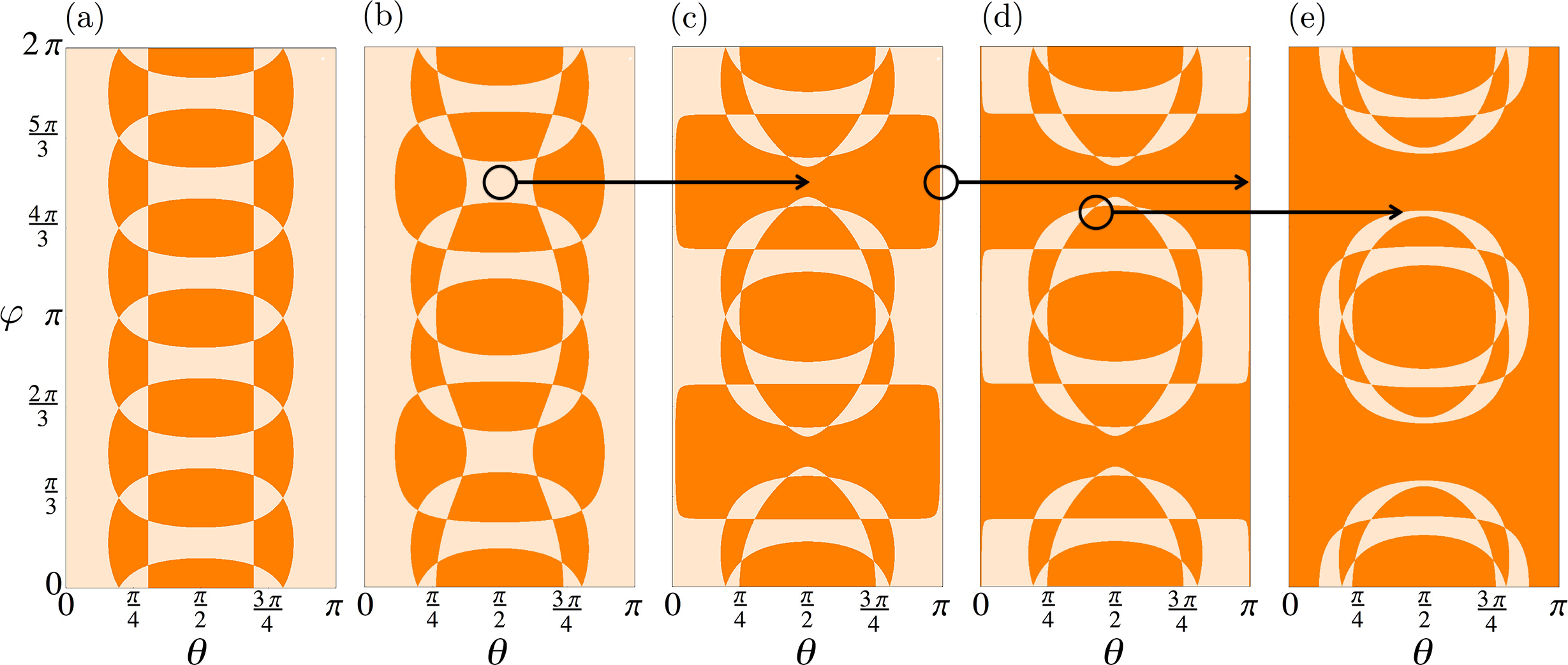}}
\caption{Phase diagrams showing the polarisation angles $(\theta, \varphi)$ at which the collective plasmon spectrum is gapped or gapless (orange and cream regions, respectively) for different 
positions of the 
$B$ sublattice, parametrised by the vector $\boldsymbol{\ell}$ (see figure~\ref{fig:deformed_lattice}). 
Panel (a) corresponds to $\boldsymbol{\ell} = (0, 0)$, i.e., a perfect honeycomb lattice (see the phase diagram in figure \ref{fig:PhaseDiagramHoneycomb}(a)), while in panels
(b) $\boldsymbol{\ell} = (0, -0.0893a)$, 
(c) $\boldsymbol{\ell} = (0, -0.1433a)$, 
(d) $\boldsymbol{\ell} = (0, -0.1437a)$ and
(e) $\boldsymbol{\ell} = (0, -0.1967a)$.}
\label{fig:PhaseTransitions}
\end{figure}

An equivalent analysis to that in section \ref{sec:PD} for the new bipartite lattices yields the resulting phase diagrams which identify the domains of LSP polarisation leading to the existence of Dirac CP quasiparticles. From equation {\eref{functionG}} we obtain the condition for gapless plasmonic dispersions at an arbitrary polarisation, 
\begin{equation}
0 \leqslant \frac{(\mathcal{C}_2^{\prime} + \mathcal{C}_3^{\prime})^2 - (\mathcal{C}_1^{\prime})^2}{4 \mathcal{C}_2^{\prime} \mathcal{C}_3^{\prime}} \leqslant 1.
\end{equation}
Figure {\ref{fig:PhaseTransitions}} shows the resulting phase diagrams in the $(\theta, \varphi)$ parameter space for five different lattices, where the cream and orange regions indicate gappless and gapped CP spectra, respectively. While figure {\ref{fig:PhaseTransitions}}(a) corresponds to the undeformed honeycomb lattice for which $\boldsymbol{\ell}=(0,0)$ (see also figure \ref{fig:PhaseDiagramHoneycomb}(a)), figures {\ref{fig:PhaseTransitions}}(b)-(e) are obtained by progressively displacing the $B$ sublattice in the $-\hat{\mathbf{y}}$ direction (see figure \ref{fig:deformed_lattice}). The phase diagram in figure \ref{fig:PhaseTransitions}(b), obtained for $\boldsymbol{\ell} = (0, -0.0893a)$
 is topologically equivalent to the one of the undeformed lattice (figure \ref{fig:PhaseTransitions}(a)). Despite the rather small displacement (less than $\unit[10]{\%}$ of $a$), the broken discrete rotational symmetry of the lattice induces an evident bulging of the phase diagram compared to the undeformed case.
 For larger deformations, in figures \ref{fig:PhaseTransitions}(b) to \ref{fig:PhaseTransitions}(e) we observe that the topology of the phase diagram changes, as indicated by the arrows in the figure. For example, when going from $\boldsymbol{\ell} = (0, -0.0893a)$ to $\boldsymbol{\ell} = (0, -0.1433a)$ (figure \ref{fig:PhaseTransitions}(b) to \ref{fig:PhaseTransitions}(c)), a compact gapless domain splits into two disconnected pockets separated by a gapped phase. In contrast, when going from $\boldsymbol{\ell} = (0, -0.1433a)$ to $\boldsymbol{\ell} = (0, -0.1437a)$ (figure \ref{fig:PhaseTransitions}(c) to \ref{fig:PhaseTransitions}(d)), the topological phase transition occurs by annihilation of a gapless domain into a gapped one. Finally, when going from $\boldsymbol{\ell} = (0, -0.1437a)$ to $\boldsymbol{\ell} = (0, -0.1967a)$ (figure \ref{fig:PhaseTransitions}(d) to \ref{fig:PhaseTransitions}(e)), the transition occurs by merging three topologically disconnected gapless domains into one. Even larger deformations lead to a phase diagram that is mostly dominated by gapped phases. 

The five phase diagrams shown in figure \ref{fig:PhaseTransitions} are snapshots from the video 
\href{http://www.ipcms.unistra.fr/wp-content/uploads/2014/11/deformed_bipartite_lattice.mp4}{\texttt{deformed\_bipartite\_lattice.mp4}}, which shows a large succession of phase diagrams corresponding to lattice deformations from 
$\boldsymbol{\ell}=(0,0)$ to $\boldsymbol{\ell}=(0, -0.2a)$.
As is clear from the video and can be inferred from panels (c) and (d) in figure {\ref{fig:PhaseTransitions}}, the spectrum for the polarisation $\theta=0$, for which the analogy between our CPs and electrons in graphene holds \cite{weick13_PRL}, is gapped for deformations larger than $\boldsymbol{\ell} \simeq (0, -0.1435a)$.
Whilst we are not discussing elastic deformations of real graphene membranes \cite{maria08_PRL, maria10_PRB, maria12_PRB} it is insightful to note that the literature on strained graphene suggests that a ca.\ $\unit[20]{\%}$ change in one of the nearest-neighbour hopping lengths is needed to merge the two Dirac points and open a gap in the spectrum \cite{monta09_PRB, monta09_EPJB, perei09_PRB}.  

\begin{figure}[tb]
\centerline{\includegraphics[width=\columnwidth, trim= 0cm 0cm 0cm 0cm, clip=true]{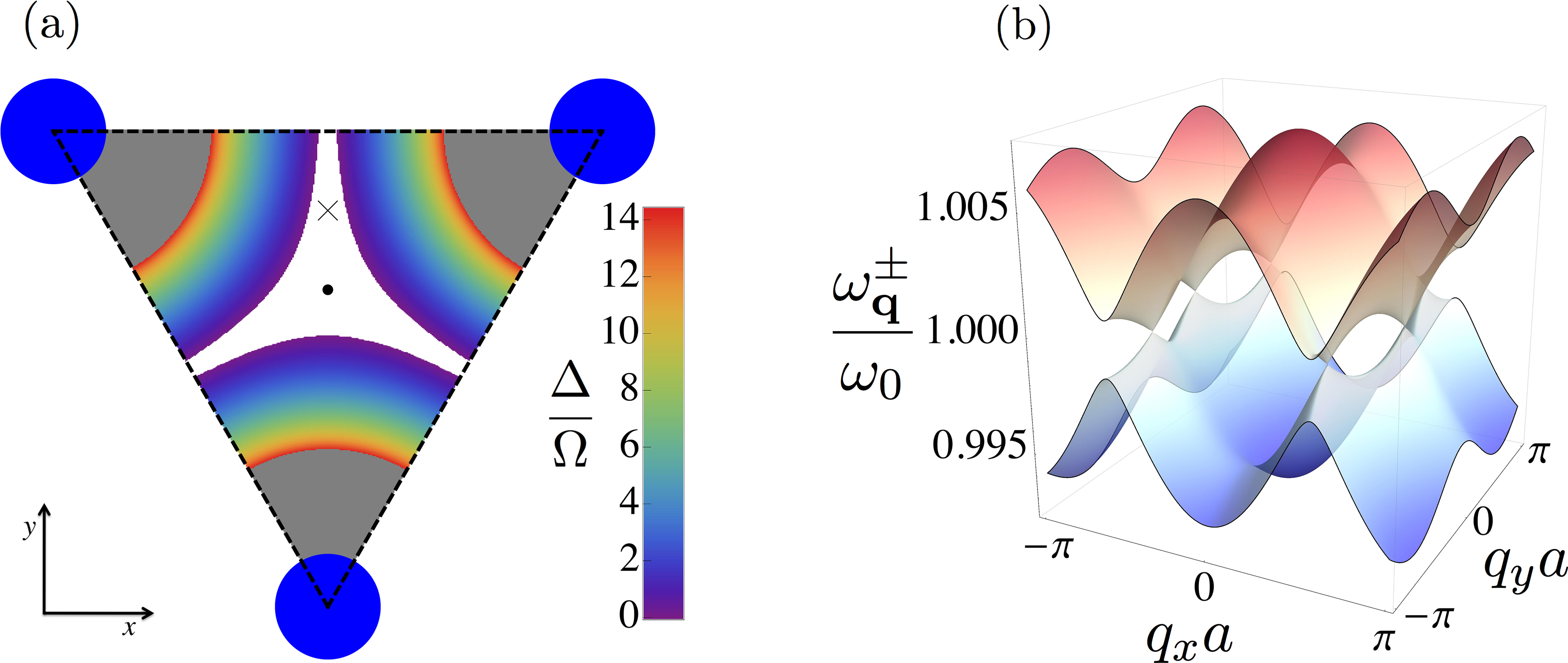}}
\caption{{(a) Band gap $\Delta$ as a function of the position of the $B$ sublattice for $\theta=0$. 
For a perfect honeycomb array, the $B$ sublattice is located at the center of the triangle (black dot) 
defined by the three neighbouring nanoparticles in sublattice $A$ (depicted as blue circles in the figure). 
The white region indicates the absence of a gap, while the colour scale gives the size of the gap in units of $\Omega$. 
The grey areas indicate regions for which the interparticle distance is smaller than $3r$ and where the dipole-dipole approximation breaks down. The cross, corresponding to a displacement $\boldsymbol{\ell}=(0, a/4)$, indicates a rather large sublattice deformation for which the spectrum, plotted in panel (b), is still characterised by the presence of massless Dirac quasiparticles (see text). 
In the figure, $a=6r$ and $\epsilon_\mathrm{m}=1$.
        }}
        \label{fig:TriangleDiagram}
\end{figure}

Figure \ref{fig:TriangleDiagram} further illuminates the qualitatively distinct physical regimes realisable in different lattices for the same dipole polarisation. For $\theta=0$, figure \ref{fig:TriangleDiagram}(a) shows the CP band gap $\Delta$ in units of the coupling $\Omega$ as a function of the position of the $B$ sublattice (white regions indicate a gapless CP dispersion). For a perfect honeycomb array, the $B$ sublattice is located at the center of the triangle (black dot) defined by the three neighbouring nanoparticles in sublattice $A$ (depicted as blue circles in the figure). The grey areas indicate regions for which the interparticle distance is smaller than $3r$ and where the dipole-dipole approximation \eref{eq:V} breaks down such that higher multipoles need to be taken into account. As can be seen from the figure, the gapless dispersion is robust in a sizeable region around the point $\boldsymbol{\ell}=(0,0)$ (see dot in the figure).
While shifting the $B$ sublattice towards the $A$ one, a non-vanishing gap forms, whose magnitude progressively increases.\footnote{Note that the discussions hitherto have considered nearest-neighbour interactions only, an approach shown to qualitatively capture the relevant physics for the perfect honeycomb structure \cite{weick13_PRL}. One must however be cautious in wielding such a model for too large values of $\boldsymbol{\ell}$ where further neighbour terms  in the interaction become increasingly important.}

To emphasise that indeed the gapless domain in figure \ref{fig:TriangleDiagram}(a) supports Dirac CPs, we focus on the particular displacement $\boldsymbol{\ell}=(0, a/4)$ indicated by a cross in the figure. For such a bipartite hexagonal lattice, the two inequivalent Dirac points are located at $\pm \mathbf{K}_{{\rm D}} = \pm \frac{2}{\sqrt{3}a} (\arccos (-\frac{\sqrt{2197}}{250}), 0)$, close to which the function $f_\q'$ defined in equation \eref{eq:f_q_prime} expands as 
$f^{\prime}_{\bf q} \simeq -\frac{96a}{125} ( \pm \sqrt{\frac{20101}{2197}} k_x + \mathrm{i} k_y )$.
Thus, close to the Dirac points the CP dispersion 
describes massless Dirac CPs with an angular-dependent group velocity, as shown in figure \ref{fig:TriangleDiagram}(b).

%=================================================================
%=================================================================
%=================================================================
%=================================================================
\section{Conclusion}
\label{sec:ccl}

We have studied the tunable plasmonic response of an artificial analogue of graphene constituted by a 2D hexagonal bipartite lattice of metallic nanoparticles. We have analysed the band structure of collective plasmons stemming from the near field dipolar coupling between localised surface plasmons in individual nanoparticles. In the case of a perfect honeycomb array, we have explored in detail the rich phase diagram of gapless and gapped phases emerging from tilting the polarisation of the dipole moments. We have shown that each topologically disconnected gapless phase supports collective plasmons that are effectively described by a massless Dirac Hamiltonian. The emergence of gapped phases from gapless ones occurs via the coalescence of Dirac cones while progressively tilting the dipole orientation, in analogy to what would occur for electrons in real graphene under extreme elastic strain. For perfect honeycomb arrays with two inequivalent sublattices we unveil that collective plasmons are effectively described as gapped Dirac particles with an energy-dependent Berry phase. 

We also studied the plasmonic properties of bipartite hexagonal arrays obtained from the honeycomb structure by rigidly shifting one sublattice with respect to the other. This system is still highly tunable with the orientation of the dipoles and supports gapless Dirac plasmons as well as gapped phases. The phase diagram undergoes significant changes for lattice displacements of a few percent, with a series of topological phase transitions that split and annihilate gapless Dirac phases. Our analysis shows that a large family of bipartite arrays of metallic nanoparticles can support Dirac plasmonic modes. These could efficiently transport highly confined radiation through the 2D metamaterial. While gapless Dirac phases exist in generic hexagonal bipartite lattices, they occupy a progressively small portion of the phase diagram for increasing deformations. This analysis will be important for designing the properties of novel plasmonic metamaterials based on bipartite lattices, e.g., by assembling dimers in ordered structures or by means of self-assembled materials with less symmetry than the honeycomb structure.

The collective plasmons in our work constitute the building block for the analysis of the optical properties of Dirac metamaterials. This would involve a quantum treatment of the coupling between photons and collective plasmons resulting in tunable plasmon polaritons that may inherit some of the effective Dirac-like properties studied in the present work. A preliminary work in this direction \cite{weick14_EPJB} has shown that plasmon polaritons in a simple 3D cubic array of metallic nanoparticles present a band structure that is tunable with the polarisation of light, leading to birefringence which is purely due to interaction effects between dipolar localised surface plasmons.

A crucial question to be addressed in the quest of designing arrays of metallic nanoparticles supporting Dirac-like collective plasmons is the inherent damping that these collective modes suffer from. Localised surface plasmons in individual metallic nanoparticles are subject to three main sources of dissipation \cite{kreibig}: absorption (Ohmic) losses, that do not depend on the size of the nanoparticle, radiation losses that scale as the volume of the nanoparticle, and Landau damping, a purely quantum-mechanical effect whose associated decay rate is inversely proportional to the nanoparticle radius. The two latter damping mechanisms suggest that there exists an optimal nanoparticle radius for which the total damping rate is minimised, estimated in reference \cite{weick13_PRL} to be around $\unit[8]{nm}$ for silver nanoparticles. The interaction between the localised surface plasmons in each nanoparticle forming the bipartite lattice may change the picture above. Indeed, recent theoretical results on metallic nanoparticle dimers \cite{brand14_preprint} suggest that while Landau damping is weakly influenced by the dipolar interaction, radiation damping strongly depends of the wavelength of the collective mode. Further work will be needed in order to investigate this important issue.

%===========================================================================
%===========================================================================
%===========================================================================
%===========================================================================
\ack
We acknowledge the CNRS PICS program (Contract No.\ 6384 APAG), the French
National Research Agency ANR (Project No.\ ANR-14-CE26-0005 Q-MetaMat), and
the Royal Society (International Exchange Grant No.\ IE140367) for financial support.

\appendix
%===========================================================================
%===========================================================================
%===========================================================================
%===========================================================================
\section{Video descriptions}
\label{sec:video}
In this appendix, we briefly describe the three videos accompanying our paper.

\subsection{Video \href{http://www.ipcms.unistra.fr/wp-content/uploads/2014/11/tunable_CP_dispersion.mp4}{\texttt{tunable\_CP\_dispersion.mp4}}}

(a) Phase diagram for the perfect honeycomb lattice showing at which polarisation angles $(\theta, \varphi)$ the collective plasmon spectrum is gapless (white regions) or gapped (coloured regions). 
The black lines indicate where one of the interaction parameters $\mathcal{C}_j$ is zero 
($\mathcal{C}_{1} = 0$, $\mathcal{C}_2=0$ and $\mathcal{C}_3=0$ along the solid, dashed and dotted lines, respectively). 
(b) Collective plasmon dispersion corresponding to the LSP polarisation $(\theta, \varphi)$ indicated by a red dot in panel (a).
The latter spans the following polarisation angles: 
\begin{enumerate}[(i)]
\item
$(0\leqslant\theta\leqslant\pi/2, \varphi=\pi/4)$,
\item
$(\theta=\pi/2, \pi/4\leqslant\varphi\leqslant\pi/2)$,
\item
$(\arcsin{(1/\sqrt{3})}\leqslant\theta\leqslant\pi/2, \varphi=\pi/2)$, 
\item
$(\theta=\arcsin{(1/\sqrt{3})}, \pi/4\leqslant\varphi\leqslant\pi/2)$.
\end{enumerate}
Coupling parameter used in the video: $\Omega/\omega_0=0.01$.

\subsection{Video \href{http://www.ipcms.unistra.fr/wp-content/uploads/2014/11/broken_inversion_symmetry.mp4}{\texttt{broken\_inversion\_symmetry.mp4}}}

Collective plasmon dispersions from equation \eref{equ:FinalDispersion} with LSP polarisation 
normal to the plane ($\theta=0$), for increasing values of $\delta\omega$, as indicated by the red arrow. 
Coupling parameter used in the video: $\tilde\Omega/\bar\omega_0=0.01$.

\subsection{Video \href{http://www.ipcms.unistra.fr/wp-content/uploads/2014/11/deformed_bipartite_lattice.mp4}{\texttt{deformed\_bipartite\_lattice.mp4}}}

(a) Phase diagram showing the polarisation angles $(\theta, \varphi)$ at which the collective plasmon
 spectrum is gapped or gapless (orange and cream regions, respectively) for different positions of the 
$B$ sublattice, parametrised by the vector $\boldsymbol{\ell}$ as indicated in the video. The corresponding deformed structure is shown in panel (e), 
where the $A$ (immobile) sublattice corresponds to the blue dots, while the $B$ (moving) sublattice corresponds to the red dots. 
Panels (b) and (c) display the same phase diagram as in panel (a) on the $(\theta, \varphi)$ unit sphere, viewed from 
(b) $(\theta=\pi/2, \varphi=0)$ and 
(c) from the north pole $\theta=0$.
(d) Corresponding collective plasmon dispersion for $\theta=0$.
In the video $\tilde\Omega/\omega_0 = 0.01$ and we are considering bipartite hexagonal lattices of 
identical nanoparticles.

%===========================================================================
%===========================================================================
%===========================================================================
%===========================================================================

\end{document}